\documentclass[11pt,onecolumn]{article}

\usepackage{graphicx}
\usepackage{epstopdf} 
\usepackage{amsmath}
\usepackage{amssymb}

\usepackage{cite}
\usepackage{color}

\usepackage{layout}

\usepackage[labelfont=bf,labelsep=period,justification=raggedright]{caption}

\makeatletter
\renewcommand{\@biblabel}[1]{\quad#1.}
\makeatother

\date{}
\usepackage{fancyhdr}
\newcommand{\D}{\mathfrak D}
\newcommand{\etal}{{\it et al.}}
\newcommand{\rmd}{{\mathrm d}}
\newcommand{\tr}{{\mathrm {tr}}}
\newcommand{\norm}{{\mathrm {norm}}}
\newcommand{\sgn}{{\mathrm {sgn}}}

\begin{document}


\title{\bf \textsf{Orthogonal Invariant Sets of the Diffusion Tensor and the Development of a Curvilinear Set Suitable for Low-Anisotropy Tissues}
\thanks{Published in PLoS ONE, November 2013. This article is reproduced here because of formatting errors by PLoS ONE in the post-production article. PLoS ONE 8(11): e78798. doi:10.1371/journal.pone.0078798}}


\author{{\bf \textsf{Robin A. Damion}}$^{1,2}$,
{\bf \textsf{Aleksandra Radjenovic}}$^{3,\dagger}$,
{\bf \textsf{Eileen Ingham}}$^2$, \\
{\bf \textsf{Zhongmin Jin}}$^{2,4}$, 
{\bf \textsf{Michael E. Ries}}$^{1,\dagger\dagger}$
\\ \\
\small $^1$ School of Physics and Astronomy, \\ 
\small University of Leeds, Leeds, West Yorkshire, UK.
\\ \\
\small $^2$ Institute of Medical and Biological Engineering, \\ 
\small School of Mechanical Engineering, University of Leeds, Leeds, West Yorkshire, UK.
\\ \\
\small $^3$ National Institute for Health Research, Leeds Musculoskeletal Biomedical Research Unit, \\
\small Chapel Allerton Hospital, University of Leeds, Leeds, West Yorkshire, UK.
\\ \\
\small $^4$ School of Mechanical Engineering, \\
\small Xi'an Jiaotong University, Xi'an, Shaanxi, People's Republic of China. 
\\ \\
\small $\dagger$ E-mail: \texttt{a.radjenovic@leeds.ac.uk}
\\
\small $\dagger\dagger$ E-mail: \texttt{m.e.ries@leeds.ac.uk}
}

\maketitle

\begin{abstract}
We develop a curvilinear invariant set of the diffusion tensor which may be applied to Diffusion Tensor Imaging measurements on tissues and porous media. This new set is an alternative to the more common invariants such as fractional anisotropy and the diffusion mode. The alternative invariant set possesses a different structure to the other known invariant sets; the second and third members of the curvilinear set measure the degree of orthotropy and oblateness/prolateness, respectively. The proposed advantage of these invariants is that they may work well in situations of low diffusion anisotropy and isotropy, as is often observed in tissues such as cartilage.

We also explore the other orthogonal invariant sets in terms of their geometry in relation to eigenvalue space; a cylindrical set, a spherical set (including fractional anisotropy and the mode), and a log-Euclidean set. These three sets have a common structure. The first invariant measures the magnitude of the diffusion, the second and third invariants capture aspects of the anisotropy; the magnitude of the anisotropy and the shape of the diffusion ellipsoid (the manner in which the anisotropy is realised). We also show a simple method to prove the orthogonality of the invariants within a set.
\end{abstract}

\section*{Introduction}
Fluids in partially-ordered biological tissues such as brain~\cite{LeBihan2012}, muscle~\cite{Sinha2011,Benson2011} and cartilage~\cite{deVisser2008_1,Filidoro2005,Meder2006} have been shown to exhibit anisotropic translational diffusion when measured with techniques such as diffusion tensor imaging (DTI)~\cite{LeBihan2001}. The fluids, which usually consist primarily of water, are inherently isotropic liquids and therefore any measured anisotropy in their translational motion is a result of the influence of a locally anisotropic environment on the motional statistics of the liquid molecules. The fact that the anisotropy is caused by the local environment is useful because it provides a source of additional contrast in magnetic resonance images and it is often possible to infer something about the local microstructure of the tissue (or porous medium) from the anisotropy of the diffusion tensor and its eigenvectors.

Normalised eigenvectors provide information about the local orientation of structures in the tissue, while the anisotropy (a comparison of the eigenvalues) provides a measure of the relative magnitudes of the eigenvectors. However, there is no single definition of anisotropy. Currently, the most common measure of diffusion anisotropy is fractional anisotropy~\cite{Basser1996} 
\begin{equation}\label{eq:FA}
A_{\rm{f}}=\sqrt{\frac{3}{2}\frac{\sum_{i=1}^3\left( \lambda_i - \bar{\lambda} \right)^2}{\sum_{i=1}^3\lambda_i^2}},
\end{equation}
(displayed in its most familiar form) where $\lambda_i$ $( i=1,2,3)$ are the three eigenvalues of the diffusion tensor, and $\bar{\lambda}=(\lambda_1+\lambda_2+\lambda_3)/3$ is the mean of the eigenvalues. Fractional anisotropy quantifies anisotropy by the spread (standard deviation) of the eigenvalues and is normalised by the quantity $\sum_i\lambda_i^2$ which provides a scale-independent, dimensionless measure. Importantly, and unlike some previous attempts to define anisotropy~\cite{Basser1996}, $A_{\rm{f}}$ is invariant. That is, the quantity is independent of the choice of coordinate system; in particular, with regard to rotations, but also more general coordinate transformations. This coordinate independence is a consequence of the application of tensor calculus which can be used to express $A_{\rm{f}}$, as was done by Basser and Pierpaoli~\cite{Basser1996}. Invariant quantities derived from tensor calculus are typically symmetric functions of the eigenvalues, as can be seen in Eq.~(\ref{eq:FA}).

Some other invariant measures that have been proposed in the literature are relative anisotropy~\cite{LeBihan2001,Basser1996}, geodesic anisotropy~\cite{Batchelor2005,Fletcher2007}, and volume ratio~\cite{LeBihan2001}. The first two of these are similar to fractional anisotropy in the sense that they are based on the spread of the eigenvalues. Volume ratio is a little different in structure and is based on the volume of the diffusion ellipsoid which is proportional to the determinant of the diffusion tensor. However, like fractional anisotropy, these quantities do not give us the full picture of the anisotropy. Except for volume ratio, these quantities calculate, in different ways, measures of the magnitude of the anisotropy but give us little idea of the shape of the diffusion ellipsoid. Volume ratio gives a single mixed measure of anisotropy magnitude and ellipsoid shape~\cite{Bahn1999}.

The shape of the diffusion ellipsoid can be generally classified by four archetypal cases or states; isotropic, orthotropic, prolate, and oblate, although only three are independent. For these, sets of eigenvalues can take the form
\begin{equation}\label{eq:cases}
\left\{ \lambda_1, \lambda_2, \lambda_3 \right\} =
\begin{cases}
\left\{ \bar{\lambda}, \bar{\lambda}, \bar{\lambda}\right\}, & \mbox{Isotropic}\\ 
\left\{ \bar{\lambda}\left( 1+\alpha \right), \bar{\lambda}, \bar{\lambda}\left( 1-\alpha \right) \right\}, & \mbox{Orthotropic}\\
\left\{ \bar{\lambda}\left( 1+2\alpha \right), \bar{\lambda}\left( 1-\alpha \right), \bar{\lambda}\left( 1-\alpha \right) \right\}, & \mbox{Prolate}\\
\left\{ \bar{\lambda}\left( 1+\alpha/2 \right), \bar{\lambda}\left( 1+\alpha/2 \right), \bar{\lambda}\left( 1-\alpha \right) \right\}, & \mbox{Oblate}
\end{cases}
\end{equation} 	 
where $\alpha\in (0,1)$. Note that these states are parameterized by $\alpha$ strictly on the open interval. That is, $\alpha=1$ should be regarded as an asymptotic limit because at this point one or more eigenvalues would be zero and the determinant of the diffusion tensor would vanish, rendering the tensor noninvertible. At the other end of the interval $\alpha=0$ and the states become a single degenerate state; the isotropic case.   

Fractional anisotropy does not easily distinguish between the states indicated above. For prolate ellipsoids, $A_{\rm{f}} \in (0,1)$, for oblate ellipsoids $A_{\rm{f}} \in (0,1/\sqrt{2})$, and for the orthotropic case $A_{\rm{f}} \in (0,\sqrt{3/5})$. Thus, only when $A_{\rm{f}} \in [\sqrt{3/5},1)$ can we identify the ellipsoid shape: prolate. In principle, the isotropic case is also identifiable as $A_{\rm{f}} = 0$  but experimental noise~\cite{Hasan2004} inevitably places a lower limit on the measured value of fractional anisotropy, meaning that $A_{\rm{f}} = 0$ is unachievable in practice.

There have been some attempts to define quantities which describe the shape of the ellipsoid. For example, Peled \etal~\cite{Peled1998} and Westin \etal~\cite{Westin2002} defined three quantities to measure the linear (prolate), planar (oblate) and spherical (isotropic) aspects of the diffusion tensor. These quantities were defined as functions on magnitude-ordered eigenvalue sets and therefore the functions treated the eigenvalues asymmetrically. The main difficulty with this approach is that averaging of magnitude-ordered eigenvalues usually introduces a statistical bias in the distribution of the eigenvalues, leading to inaccurate measurements of anisotropy~\cite{Basser2000,Pierpaoli1996}.

A solution to the problem of finding invariant measures of anisotropy which encode the shape of the diffusion ellipsoid, and which do not depend on having to magnitude-order the eigenvalues, was proposed by Ennis and Kindlmann~\cite{Ennis2006}. These authors, basing their work on that of Criscione \etal~\cite{Criscione2000}, showed that there exist orthogonal sets (triplets) of tensor invariants which can be applied to the diffusion tensor to give useful measures of anisotropy. They found two sets of invariants which are related to cylindrical and spherical coordinates in eigenvalue space. The latter set was shown to contain fractional anisotropy as a measure of the magnitude of anisotropy, and an invariant known as the mode; a measure of the shape of the diffusion ellipsoid.

In the next section of this paper we review the orthogonal cylindrical and spherical sets using a more geometric approach than was used by Ennis and Kindlmann. These authors gave a proof of the orthogonality of these measures in the same way as Criscione \etal ~but this proof is based on tensors in the full six-dimensional space of the diffusion tensor and the connection with the coordinate system of the eigenvalue space is obscured. Here, we give an alternative proof of the orthogonality of these invariant measures which maintains a direct connection with the eigenvalue coordinates and therefore presents a clearer picture of the geometry and the meaning of the invariants. In fact the geometrical, coordinate-based approach we take here is similar to that used by Bahn~\cite{Bahn1999} prior to the work of the abovementioned authors. Bahn was able to show how the spherical coordinates of eigenvalue space were invariant themselves or related to the invariants forming an orthogonal set. What he did not show was how these invariants were related to invariants of the full diffusion tensor and also explicitly how his skewness parameter (related to the mode) determined the shape of the ellipsoid, nor did he fully explore the invariants relating to the cylindrical set. 

We also look at two further orthogonal sets. One set is related to recent work using the log-Euclidean metric for averaging and interpolation of the diffusion tensor~\cite{Batchelor2005,Fletcher2007}. The final set, the curvilinear set, has a different structure to the previous three and makes the orthotropic ellipsoid case more apparent. The incentive behind the definition of the latter set was our discovery that the third invariants (measuring the mode or its equivalent) of the previous three sets (spherical, cylindrical, log-Euclidean) do not function well for diffusion exhibiting isotropy or a low degree of anisotropy. 

The primary purpose of this paper is to introduce the curvilinear invariant set and relate it and compare it mathematically to the other invariant sets. It is not our intention to carry out a numerical or experimental validation of the curvilinear invariants here but to present their definition, mathematical properties, and to propose them as potentially useful invariants which warrant further study.

\section*{Theory}
\subsection*{Orthogonality and the Metric}
To prove the orthogonality of invariants, we are going to treat them as if they were the coordinates on some manifold $\mathcal{M}$ and use the metric tensor $g_{ij}$, where $i,j\in \{1,2,\dots, n\}$, and $n$ is the dimension of the manifold. The metric is a symmetric tensor which is used to calculate the distance along a specified path on a manifold. Here, we shall only be concerned with infinitesimal distances $\rmd s$ given by (using the Einstein summation convention)
\begin{equation}\label{eq:metric}
\rmd s^2 = g_{ij} \rmd x^i\rmd x^j,
\end{equation}
where $\rmd x^i$ are infinitesimal changes in the coordinates $x^i \in \mathcal{M}$. This is merely an alternative way of displaying the metric tensor. For our purposes, the significance of using the metric is that for real orthogonal coordinates the metric is diagonal and hence there will be only terms of the form $g_{ii}(\rmd x^i)^2$ and no terms of the form $g_{ij}\rmd x^i\rmd x^j$ with $i\not= j$.

Let $\mu:\mathcal{M}\to \mathcal{N}$ be a map from manifold $\mathcal{M}$ to $\mathcal{N}$ (which, for simplicity, we will assume to be of the same dimension as $\mathcal{M}$) such that $\mu:x^i\mapsto X^i\in\mathcal{N}$. If  $\mathcal{N}$ is the same manifold as $\mathcal{M}$, $\mu:\mathcal{M}\to \mathcal{M}$ is merely a coordinate transformation on $\mathcal{M}$. To calculate the new metric $\gamma_{ij}$ on $\mathcal{N}$, we need the Jacobian matrix $\mathbf J$ such that $J^i_j=\partial x^i/\partial X^j$. The new metric and the orthogonality can then be assessed by substituting $\rmd x^i = J^i_j \rmd X^j$ into Eq.~(\ref{eq:metric}) or by computing the new metric directly by $\mathbf\gamma = \mathbf J^{\rm T} \mathbf g \mathbf J$, where $\mathbf J^{\rm T}$ is the transpose of the Jacobian matrix. In index notation $\gamma_{ij}=g_{kl}J^k_iJ^l_j$.

Although coordinates are usually designated with superscript indices, as we have employed in the above two paragraphs, below for convenience we will use subscripted indices to avoid confusion between indices and exponents. This should not lead to any ambiguity since, from here on, index notation will not be used extensively.

\subsection*{Cartesian Coordinates}
Since tensor invariants are scalar functions of eigenvalues only, we will consider the three-dimensional eigenvalue space only. We begin with the eigenvalues as orthogonal coordinates in three-dimensional Euclidean space $\mathcal{E}$. Strictly, since the eigenvalues must be positive real numbers, they reside only in the open subset that is one octant of the full Euclidean space such that $\lambda_1, \lambda_2, \lambda_3 >0$. The Euclidean metric tensor is given by 
\begin{equation}\label{eq:euclidean-tensor}
{\mathbf g} = \begin{bmatrix} 1 & 0 & 0 \\ 0 & 1 & 0 \\ 0 & 0 & 1 \end{bmatrix},
\end{equation}
and the infinitesimal distance element is given by
\begin{equation}\label{eq:euclidean-metric}
\rmd s^2 = g_{ij}\rmd\lambda^i\rmd\lambda^j = \rmd \lambda_1^2 + \rmd \lambda_2^2 + \rmd \lambda_3^2.
\end{equation}

However, it will be convenient to use a slightly different coordinate system defined, for example, by
\begin{equation}\label{eq:rotation}
\begin{bmatrix} x \\ y \\ z \end{bmatrix} = \begin{bmatrix} 1/\sqrt{6} & 1/\sqrt{6} & -\sqrt{2/3} \\ -1/\sqrt{2} & 1/\sqrt{2} & 0 \\ 1/\sqrt{3} & 1/\sqrt{3} & 1/\sqrt{3} \end{bmatrix} \begin{bmatrix} \lambda_1 \\ \lambda_2 \\ \lambda_3 \end{bmatrix},	
\end{equation}
where the matrix in this equation is an orthogonal matrix (a rotation matrix similar to that used by Bahn~\cite{Bahn1999}) and has been chosen because it preserves the orthogonality of the coordinates while defining the $z$-direction to be proportional to the tensor trace. The component of the rotation about the $z$-axis, however, is arbitrary. The matrix shown in Eq.~(\ref{eq:rotation}) is the inverse Jacobian matrix, as we have defined it above, and since in this case it is also an orthogonal matrix, $\mathbf J^{-1}=\mathbf J^{\rm T}$.

The convenience of this choice is that in the case of isotropy, when the eigenvalues are identical, $x=y=0$ and therefore isotropic tensors will be represented by points on the $z$-axis and the anisotropy is dependent on the $x$ and $y$ coordinates only. Note that while the $z$-coordinate, being proportional to the tensor trace, is an invariant, the $x$   and $y$ coordinates are not invariants and therefore we will need either to find another coordinate system or appropriate functions of the coordinates. However, this Cartesian system will serve as an important first step in establishing the geometry.

In defining our new Cartesian coordinates, we used a rotation matrix. Therefore, we know that the new coordinates will also be orthogonal and the standard Euclidean metric will be preserved. Calculating the new metric, 
\begin{equation}\label{eq:cartesian-metric}
\rmd s^2 = \rmd x^2 + \rmd y^2 + \rmd z^2,
\end{equation}
we verify the orthogonality of these coordinates by the absence of off-diagonal terms such as $\rmd x\rmd y$.

\subsection*{Cylindrical Coordinates and Invariants}
Let us change to a cylindrical polar coordinate system $\{ z, \rho, \theta\}$ such that the $z$-axis remains unchanged and
\begin{equation}\label{eq:cylindrical-coords} 
\begin{array}{rcl}
x &=& \rho \cos \theta, \\
y &=& \rho \sin \theta,
\end{array}
\end{equation}
and the metric becomes 
\begin{equation}\label{eq:cylindrical-metric}
\rmd s^2 = \rmd z^2 + \rmd \rho^2 + \rho^2\rmd \theta^2.
\end{equation}

The cylindrical invariant set~\cite{Ennis2006} can now be defined in terms of these coordinates by
\begin{eqnarray}\label{eq:cylindrical-set} 
K_1 =&  \sqrt{3}z &=  \tr (\mathbf D), \nonumber\\ 
K_2 =&  \rho &=  \norm (\tilde{\mathbf D}), \\ 
K_3 =&  -\cos 3\theta &=  3\sqrt{6}\frac{\det (\tilde{\mathbf D})}{\norm (\tilde{\mathbf D})^3},  \nonumber
\end{eqnarray}
where $\mathbf D$ is the diffusion tensor, $\tilde{\mathbf D}=\mathbf D - \frac{1}{3}\tr (\mathbf D) \mathbf I$ is the deviatoric tensor, $\mathbf I$ is the identity tensor, $\tr$ is the trace, $\det$ is the determinant, and $\norm (\mathbf A)^2 = \tr (\mathbf A^2)$ for an arbitrary tensor $\mathbf A$. The trace, norm and determinant are tensor invariants and thus Eq.~(\ref{eq:cylindrical-set}) can be defined directly in terms of the eigenvalues of $\mathbf D$. In terms of the eigenvalues, the basic tensor invariants used here and below are
\begin{equation}\label{eq:invariant-defs} 
\begin{array}{rcl}
\tr (\mathbf D) &=& \sum_{i=1}^3 \lambda_i, \\ 
\norm (\mathbf D) &=& \sqrt{\sum_{i=1}^3 \lambda_i^2}, \\ 
\norm (\tilde{\mathbf D}) &=& \sqrt{\sum_{i=1}^3 (\lambda_i-\bar{\lambda})^2}, \\ 
\det (\tilde{\mathbf D}) &=& \prod_{i=1}^3 (\lambda_i-\bar{\lambda}),
\end{array}
\end{equation}
(of which only three are independent) and the relations in (\ref{eq:cylindrical-set}) between the cylindrical invariants  and the cylindrical coordinates can be verified using (\ref{eq:rotation}), (\ref{eq:cylindrical-coords}), and (\ref{eq:invariant-defs}).

The cylindrical invariants have the following interpretation. $K_1\in (0,\infty)$ is proportional to the tensor trace and is a measure of the magnitude of the diffusion tensor. $K_2$ and $K_3$, being functions on the plane perpendicular to the $z$-axis, quantify the anisotropy. Like fractional anisotropy, $K_2\in (0,\infty)$ measures the magnitude of the anisotropy by the spread of the eigenvalues. However, unlike fractional anisotropy, $K_2$ is not normalised and therefore not scale-independent. In the cylindrical coordinate system $K_2$ has the direct interpretation of being the perpendicular distance between point $(z, \rho, \theta)$ and the $z$-axis; {\it i.e.} the radius $\rho$. $K_3\in [-1,1]$ is known as the mode and quantifies the shape of the diffusion ellipsoid. For the three ellipsoid shapes given by (\ref{eq:cases}), $K_3=1,0,-1$ for the prolate, orthotropic and oblate cases respectively. In the case of isotropic ellipsoids, a little care needs to be taken because $K_3$ is not defined and depends on how this point (the $z$-axis) is approached.

For example, let us parameterize the prolate and oblate cases with a single parameter $\beta\in (-\frac{1}{2},1)$ via 
\begin{equation}\label{eq:cylindrical-beta}
\{\lambda_1, \lambda_2, \lambda_3 \} = \left\{ \bar{\lambda}\left( 1+2\beta \right), \bar{\lambda}\left( 1-\beta \right), \bar{\lambda}\left( 1-\beta \right) \right\},
\end{equation}
where $\beta\in (-\frac{1}{2},0)$ corresponds to oblate ellipsoids and $\beta\in (0,1)$ corresponds to the prolate case. Calculating the mode as a function of $\beta$, $K_3=\beta^3/|\beta|^3 = \sgn (\beta)$, and it is clear that a discontinuous jump occurs as we pass through the $z$-axis ($\beta=0$). A similar discontinuous jump occurs along any radial path through the $z$-axis except for those angles, $\theta$, which correspond to the orthotropic ellipsoids. In these cases, the mode is zero along the whole radial path.

The invariant set~(\ref{eq:cylindrical-set}) can be regarded as a triplet of scalar functions on the eigenvalue space with the metric~(\ref{eq:cylindrical-metric}) but it is also possible to regard these functions as coordinates on a related manifold $\mathcal{K}$. Because the function $K_3:\theta\to -\cos 3\theta$ is a six-to-one mapping, the eigenvalue manifold is a six-fold covering of the manifold associated with coordinates $\{K_i\}$. This is a consequence of the invariance of the $\{K_i\}$ which treats the eigenvalues symmetrically; there are six equivalent permutations of $(\lambda_1, \lambda_2, \lambda_3)$. The metric on $\mathcal{K}$ is
\begin{equation}\label{eq:K-metric}	 
\rmd s^2 = \frac{1}{3}\rmd K_1^{\;2} + \rmd K_2^{\;2} + \frac{K_2^{\;2}}{9(1-K_3^{\;2})}\rmd K_3^{\;2}.
\end{equation}
Again, the absence of off-diagonal terms $\rmd K_i\rmd K_j$ verifies the orthogonality of the coordinates and hence the set $\{K_i\}$ are orthogonal tensor invariants. We also note that despite the singularity in the metric on the boundary $K_3=\pm 1$, the metric is completely flat. That is, the Ricci scalar curvature vanishes everywhere.

\subsection*{Spherical Coordinates and Invariants}
The invariant $K_2$ is similar to fractional anisotropy but lacks a normalisation which makes it scale-independent. Ennis and Kindlmann~\cite{Ennis2006} showed that it is possible to modify the set to include fractional anisotropy and that this new set is closely related to spherical polar coordinates:
\begin{eqnarray}\label{eq:spherical-coords}
x & = & r \sin \phi \cos \theta,  \nonumber\\
y & = & r \sin \phi \sin \theta,  \\
z & = & r \cos \phi.  \nonumber	 
\end{eqnarray}
The metric in these coordinates is
\begin{equation}\label{eq:spherical-metric}
\rmd s^2 = \rmd r^2 + r^2\rmd\phi^2 + r^2\sin^2\phi\rmd \theta^2.
\end{equation}
The invariants associated with these coordinates are defined as~\cite{Ennis2006}
\begin{eqnarray}\label{eq:spherical-set} 
R_1 =&  r &=  \norm (\mathbf D), \nonumber\\ 
R_2 =&  \sqrt{\frac{3}{2}}\sin\phi &=  \sqrt{\frac{3}{2}} \frac{\norm (\tilde{\mathbf D})}{\norm (\mathbf D)}, \\ 
R_3 =&  -\cos 3\theta &=  3\sqrt{6}\ \frac{\det (\tilde{\mathbf D})}{\norm (\tilde{\mathbf D})^3},  \nonumber
\end{eqnarray}
where, using the definition of the norms in Eq.~(\ref{eq:invariant-defs}), one can see that $R_2$ is equivalent to the usual definition of fractional anisotropy (see Eq.~(\ref{eq:FA})).

The structure of these spherical invariants is very similar to the cylindrical invariants. $R_1$, like $K_1$, is a measure of the magnitude of the diffusion tensor but is now the radial coordinate, $r$, which corresponds to the norm of the diffusion tensor. Again, $R_2$ and $R_3$ measure the anisotropy of the tensor. $R_2$, as we have already mentioned, is exactly equivalent to fractional anisotropy, $A_{\rm f}$, and is related to the polar angle, $\phi$. It is also similar to $K_2$ in its geometrical interpretation in that it is also (proportional to) the perpendicular distance from a point $(r, \phi, \theta)$ to the $z$-axis ($\phi=0$ direction) but normalised by the radius, $r$. $R_3$ is exactly the same as $K_3$, the mode, and thus quantifies the diffusion ellipsoid shape. As with the cylindrical coordinates, it is also related to the azimuthal angle, $\theta$.

As with the cylindrical set, we can regard $\{R_i\}$ as coordinates on a related manifold $\mathcal{R}$ and demonstrate the orthogonality of these invariants by showing that the metric is purely diagonal;
\begin{equation}\label{eq:R-metric}	 
\rmd s^2 = \rmd R_1^{\;2} + \frac{R_1^{\;2}}{\frac{3}{2}-R_2^{\;2}}\rmd R_2^{\;2} + \frac{2R_1^{\;2}R_2^{\;2}}{27(1-R_3^{\;2})}\rmd R_3^{\;2}.
\end{equation}

\subsection*{Log-Euclidean Coordinates and Invariants}
Recently in the literature there has been some debate over the way in which diffusion tensors should be averaged and interpolated~\cite{Pasternak2010}. The simplest way to average tensors is to treat them as one would treat real numbers by directly adding them and dividing by the number of tensors. A consequence of this (Euclidean-averaging) method is that if the original tensors had the same traces then the resulting tensor also has this trace. There is then a natural affinity between this method and the invariant set $\{K_i\}$ since $K_1$ (the trace) is preserved in such circumstances. This is because the condition that $K_1$ is constant defines a flat plane in the eigenvalue space (with its assumed Euclidean metric) and geodesics between points in this plane are straight lines and therefore will remain in the plane.

The set $\{R_i\}$ is also consistent with Euclidean averaging even though none of the set members are generally preserved in the sense that $K_1$ is. This is because this set is also derived from the assumption of a Euclidean metric on the eigenvalue space. However, if one were interested in preserving $R_1$ in the same way that $K_1$ is preserved in the cylindrical set, then one would need to average over the surface of a sphere. As far as we are aware, no such averaging method has been proposed in the literature but one such method would be to simply average the square of the diffusion tensors and find the appropriate square-root. This method would preserve the tensor norm for tensors of equal norm.

An alternative method of averaging which has been discussed and explored in the literature is the log-Euclidean method~\cite{Batchelor2005,Fletcher2007,Arsigny2006}. This is similar to the Euclidean method except that the logarithm of the tensors is averaged. This method preserves the tensor determinant. It would therefore be natural to ask whether there exists an invariant set $\{L_i\}$ such that $L_1$ is related to the determinant. Here we show that such a set does exist and had already been found by Criscione \etal~\cite{Criscione2000}.

We begin with Eq.~(\ref{eq:rotation}) but insert the map
\begin{equation}\label{eq:log-map}
\ell_\kappa:(\lambda_1,\lambda_2,\lambda_3)\mapsto (\ln \kappa\lambda_1,\ln \kappa\lambda_2,\ln \kappa\lambda_3),
\end{equation}
 where $\kappa >0$ is an arbitrary real parameter with dimensions s/m$^2$, prior to the application of the rotation matrix. $\ell_\kappa$ maps the octant of eigenvalue space to the full Euclidean space. We now assume that the metric of Eq.~(\ref{eq:cartesian-metric}) is valid. This is now our starting point, not the metric of Eq.~(\ref{eq:euclidean-metric}), which is no longer valid.

We now follow the construction of the cylindrical invariant set equations (\ref{eq:cylindrical-coords}) to (\ref{eq:cylindrical-set}) except that now we have
\begin{eqnarray}\label{eq:log-euclidean-set} 
L_1 =&  \sqrt{3}z &=  \tr (\Lambda), \nonumber\\ 
L_2 =&  \rho &=  \norm (\tilde{\Lambda}), \\ 
L_3 =&  -\cos 3\theta &=  3\sqrt{6}\frac{\det (\tilde{\Lambda})}{\norm (\tilde{\Lambda})^3},  \nonumber
\end{eqnarray}
where $\Lambda = \ln \kappa {\mathbf D}$ and $\tilde{\Lambda}=\Lambda - \frac{1}{3}\tr (\Lambda) \mathbf I$. By using the fact that $\tr (\Lambda) = \ln (\kappa^3 \det (\mathbf D))$ and $\tilde{\Lambda} = \ln\left[ \mathbf D/\det (\mathbf D)^{1/3}\right]$ we can re-express the above as
\begin{eqnarray}\label{eq:log-euclidean-eigen-set} 
L_1 &=&  \ln (\kappa^3\D), \nonumber\\ 
L_2 &=&  \sqrt{\sum_{i=1}^3 \left[ \ln \left( \frac{\lambda_i}{\D^{1/3}}\right) \right]^2}, \\ 
L_3 &=&  \frac{3\sqrt{6}\prod_{i=1}^3 \ln \left( \dfrac{\lambda_i}{\D^{1/3}}\right)}{L_2^{\;3}},  \nonumber
\end{eqnarray}
where ${\lambda_i}$ are the eigenvalues of $\mathbf D$ and $\D = \det (\mathbf D) = \lambda_1\lambda_2\lambda_3$.

As we remarked earlier, the set of invariants $\{L_i\}$ is identical to that found in the paper by Criscione \etal~\cite{Criscione2000} (except for the inclusion of the parameter $\kappa$). The cylindrical invariants $\{K_i\}$  of Ennis and Kindlmann~\cite{Ennis2006} were derived from the invariants of Criscione \etal ~but the latter authors' log-Euclidean invariants were not mentioned in the former authors' paper. Furthermore, $L_2$ was discovered independently by Batchelor \etal~\cite{Batchelor2005} and Fletcher and Joshi~\cite{Fletcher2007} from their investigations of the space of diffusion tensors. Both sets of authors referred to $L_2$ as geodesic anisotropy.

The structure of these log-Euclidean invariants is similar to the cylindrical and spherical sets. $L_1\in (-\infty,\infty)$ is the measure of overall magnitude of the diffusion tensor. $L_2\in [0,\infty)$ is similar to fractional anisotropy in these log-Euclidean coordinates and has a similar interpretation, being the perpendicular distance, $\rho$, between the point $(z,\rho,\theta)$ and the $z$-axis (isotropy). Since this is the distance of a path which is also a geodesic in this space (a straight line in Euclidean space), $L_2$ was called geodesic anisotropy~\cite{Batchelor2005,Fletcher2007}. The same appellation could equally be applied to $K_2$, however. A difference worth noting between $K_2$ and $L_2$ is that the latter is scale-independent, as is evidenced by the absence of the parameter $\kappa$.

$L_3\in [-1,1]$ plays an identical role to the mode as defined by $K_3$ (and $R_3$). However, since we have now assumed a different metric on the eigenvalue space, the parameterised description of the four archetypal cases is now different to that given in~(\ref{eq:cases}) and can be given by
\begin{equation}\label{eq:log-euclidean-cases}
\left\{ \lambda_1, \lambda_2, \lambda_3 \right\} =
\begin{cases}
\left\{ \D^{1/3}, \D^{1/3}, \D^{1/3} \right\}, & \mbox{Isotropic}\\ 
\left\{ \D^{1/3}e^{\alpha}, \D^{1/3}, \D^{1/3}e^{-\alpha} \right\}, & \mbox{Orthotropic}\\
\left\{ \D^{1/3}e^{2\alpha}, \D^{1/3}e^{-\alpha}, \D^{1/3}e^{-\alpha} \right\}, & \mbox{Prolate}\\
\left\{ \D^{1/3}e^{\alpha}, \D^{1/3}e^{\alpha}, \D^{1/3}e^{-2\alpha} \right\}, & \mbox{Oblate}
\end{cases}
\end{equation} 
where $\alpha\in (0,\infty)$. Alternatively, as we did previously in~(\ref{eq:cylindrical-beta}), the prolate and oblate cases can be parameterized together as
\begin{equation}\label{eq:log-euclidean-beta}
\{\lambda_1, \lambda_2, \lambda_3 \} = \left\{ \D^{1/3}e^{2\beta}, \D^{1/3}e^{-\beta}, \D^{1/3}e^{-\beta} \right\},
\end{equation}
where $\beta\in (-\infty,\infty)$  can be broken into a union of three intervals as $\beta\in (-\infty,0)\cup [0]\cup (0,\infty)$. These first and last intervals correspond to the oblate and prolate cases respectively and the point $\beta=0$  is the isotropic case. Using this parameterisation, $L_3=\sgn (\beta )$. As with $K_3$, a discontinuity exists as  $\beta$ passes from negative to positive and  $L_3$ at $\beta=0$ (isotropy) is undefined. For the orthotropic case, $L_3=0$  for all  $\alpha$ including the isotropic case, {\it i.e.} $\alpha\in [0,\infty)$.

The orthogonality of these invariants is guaranteed by the fact that they have exactly the same relation to the cylindrical coordinates as the $\{K_i\}$. Therefore, the metric is exactly analogous to Eq.~(\ref{eq:K-metric}).

\subsection*{Conformal Maps, Curvilinear Coordinates and Invariants}
In the previous sections, the orthogonal invariant sets have a common structure; the first invariant measures the magnitude of the diffusion tensor; the second invariant measures the magnitude of the anisotropy and the third invariant provides a measure of the shape of the ellipsoid. In this final section, we will describe an invariant set with a slightly different structure.

We begin by recalling that the invariant sets $\{L_i\}$ and $\{K_i\}$ were based on an underlying cylindrical coordinate system $\{z,\rho,\theta\}$ and that the second and third members of these sets are related to the plane of the polar coordinates $\{ \rho,\theta\}$. This plane is naturally regarded as the two-dimensional space of real numbers $\mathbb R^2$ but can equally be viewed as the complex plane $\mathbb C$ and, in this context, there are a family of transformations which preserve angles; the conformal maps~\cite{Jeffrey1986}.

We therefore seek conformal transformations of these coordinates which produce alternative invariant measures $\{C_i\}$ (other than those in Eqs.~(\ref{eq:cylindrical-set}) or (\ref{eq:log-euclidean-set})) such that $C_1=K_1$ or $L_1$.

Defining the complex number $Z=x+\imath y=\rho e^{\imath\theta}$, we seek conformal maps $F:\mathbb C\to \mathbb C$ such that $W=F(Z)=U(x,y)+\imath V(x,y)$. Such conformal maps are complex analytic functions with non-vanishing complex derivative (except at critical points) and guarantee the preservation of orthogonality via the Cauchy-Riemann equations
\begin{equation}\label{eq:Cauchy-Riemann}
\frac{\partial U}{\partial x} = \frac{\partial V}{\partial y} \mbox{  \rm and  } \frac{\partial U}{\partial y} = -\frac{\partial V}{\partial x}.	 
\end{equation}
Potential invariants can then be constructed from  $(C_2,C_3)=\left( f_2(U),f_3(V)\right)$, where $f_2$ and $f_3$ are differentiable functions.

The simplest maps to provide invariants appear to be $F_n:Z\mapsto Z^{3n}$ for integers $n\geq 1$, with more complicated maps being constructed from convergent polynomials or fractional linear transformations thereof. Using the simplest, non-trivial map $F_1$, with an additional $\pi/2$ rotation, we make the following assignments;
\begin{equation}\label{eq:assignments} 
\begin{array}{rrl}
C_2 =&  \rho^3\sin 3\theta, & \\ 
C_3 =& -\rho^3\cos 3\theta. & 
\end{array}
\end{equation}

At this stage, the invariants $C_2$ and $C_3$ are expressed in terms of polar coordinates and we could choose to relate them to $\{K_i\}$ or $\{L_i\}$. From this point on, we will relate them to the log-Euclidean set since the resulting invariants will then be scale-independent.

The rotation used in obtaining the invariants above conveniently introduces the minus sign for $C_3$ so that, comparing it to the mode $L_3$ (Eq.~(\ref{eq:log-euclidean-set})), we see that $C_3\propto L_3$ and $C_2\propto\sqrt{1-L_3^{\;2}}$. These two invariants can therefore be re-expressed as
\begin{equation}\label{eq:C-to-L} 
\begin{array}{rcl}
C_2 &=& L_2^{\;3}\sqrt{1-L_3^{\;2}}, \\ 
C_3 &=& L_2^{\;3}L_3. 
\end{array}
\end{equation}

The metric for coordinates $\{C_i\}$ on the related manifold $\mathcal{C}$ is
\begin{equation}\label{eq:C-metric}
\rmd s^2 = \frac{1}{3}\rmd C_1^{\;2} + \frac{\rmd C_2^{\;2}+\rmd C_3^{\;2}}{9\left( C_2^{\;2}+C_3^{\;2}\right)^{2/3}}.	 
\end{equation}
The invariants $\{C_i\}$ therefore form an orthogonal set and, as mentioned above, these invariants possess a slightly different structure to the previous three sets $\{K_i\}$, $\{R_i\}$, and $\{L_i\}$. In the previous sets, the second invariant ($i=2$) measures the magnitude of anisotropy whereas the third invariant ($i=3$) measures the shape of the ellipsoid. As shown above in Eq.~(\ref{eq:C-to-L}), $C_2$ and $C_3$ are both functions of $L_2$ and $L_3$; both contain information about the magnitude of the anisotropy and the ellipsoid shape but $C_2$ and $C_3$ contain different information about the shape.  Table~\ref{tab:C-and-R} shows a comparison between the spherical measures, $R_2$ and $R_3$, and the curvilinear measures, $C_2$ and $C_3$. It can be seen that $C_3$ is similar to $R_3$ except that the magnitude of $C_3$ is modulated by a positive function of  $L_2$. Therefore, $C_3$ measures the degree of prolateness or oblateness of the ellipsoid, the two cases being differentiated by the sign of $C_3$. $C_2$, on the other hand, measures the degree of orthotropy. This is in contrast to the spherical measures in which orthotropy is inferred from the absence of prolateness or oblateness; $R_3=0$ but $R_2\neq 0$.

We also note that the discontinuity in $L_3$ and $R_3$ as one passes through the isotropic state is absent from $C_2$ and $C_3$. If the parameterisation of Eq.~(\ref{eq:log-euclidean-beta}) is used for the prolate and oblate cases, the pair $(C_2,C_3)=(0,6^{3/2}\beta^3)$ and it is clear that smooth behaviour is exhibited for all values of $\beta$. Likewise, using the parameterisation for the orthotropic case in Eq.~(\ref{eq:log-euclidean-cases}), $(C_2,C_3)=(2^{3/2}|\alpha |^3,0)$, which is also smooth (up to second derivative) for all values of $\alpha$, even if we extend the parameterisation to negative values; $\alpha\in(-\infty,\infty)$.  

To calculate the curvilinear measures, we can first calculate the log-Euclidean measures using Eq.~(\ref{eq:log-euclidean-set}) or~(\ref{eq:log-euclidean-eigen-set}) and use~(\ref{eq:C-to-L}) together with $C_1=L_1$, but it is also interesting and useful to show how they can be calculated directly from the diffusion and deviatoric tensors;
\begin{eqnarray}\label{eq:curvilinear-set} 
C_1 &=&  \tr (\Lambda), \nonumber\\ 
C_2 &=&  \sqrt{\norm (\tilde{\Lambda})^6 - 54\;\det (\tilde{\Lambda})^2}, \\ 
C_3 &=&  3\sqrt{6}\;\det (\tilde{\Lambda}).  \nonumber
\end{eqnarray}

Finally, we note that in constructing the curvilinear set of invariants $\{C_i\}$, we have done so using the invariant sets related to cylindrical coordinates on the eigenvalue space. This was convenient because the polar coordinates define a plane on which we could employ the conformal transformations. However, something similar can be done with the spherical set $\{R_i\}$. In this case, a stereographic projection can be used to conformally map the spherical coordinates $\{\phi,\theta \}$ to the complex plane and then invariants can be defined similarly to those in this section.

\begin{table}[!t]
\center
\caption{
\bf{Spherical and curvilinear anisotropies}}
\begin{tabular}{c|r|r|r|r}
\bf Ellipsoid shape & $R_2$ & $R_3$ & $C_2$ & $C_3$  \\ 
\hline
Isotropic       &  $0$  &  $0$  &  $0$  &  $0$   \\
Orthotropic     &  $>0$ &  $0$  &  $>0$ &  $0$   \\
Prolate         &  $>0$ &  $1$  &  $0$  &  $>0$  \\
Oblate          &  $>0$ &  $-1$ &  $0$  &  $<0$  \\
\end{tabular}
\begin{flushleft}Comparison of the spherical $(R_2, R_3)$ and curvilinear measures $(C_2, C_3)$ of anisotropy for the four archetypical ellipsoid shapes.
\end{flushleft}
\label{tab:C-and-R}
\end{table}

\section*{Discussion}
Up to permutations of the eigenvalues, any invariant set can be inverted to return the eigenvalues (see Criscione \etal~\cite{Criscione2000}). Therefore any two invariant sets  contain the same information. However, different sets display this information in different ways and can make certain aspects of this information less or more clear.

As described in the Theory section, the sets $\{K_i\}$, $\{R_i\}$ and $\{L_i\}$ are all similar in structure. What distinguishes the set $\{C_i\}$ is that it makes orthotropic anisotropy more explicit. In the previous three sets, orthotropy is inferred from the combination of, for example, $R_2>0$ and $R_3=0$, whereas $C_2$ is a more direct measure of the degree of orthotropy.  That is the single invariant $C_2$ provides a much clearer picture of this aspect of the anisotropy.

Whilst it could be seen as a disadvantage that the invariants $C_2$ and $C_3$ have a cubic non-linearity (see equation~(\ref{eq:C-to-L})), it is this property that has removed the discontinuity of the mode at isotropy. This feature of $C_2$ and $C_3$ may prove to be advantageous when dealing with noisy diffusion tensors that are close to isotropic. Since the mode, $R_3$, possesses a discontinuity at isotropy, small fluctuations in the eigenvalues can produce large fluctuations in $R_3$. On the other hand, since $C_2$ and $C_3$ behave smoothly at isotropy it is expected that the effects of noise will be decreased in the region of isotropic tensors.

\section*{Limitations of the Study, Open Questions and Future Work}
In the current work, we have focused on the mathematical development of the curvilinear invariants, their properties and their relation to other invariants. Our current interest is in the application of the new invariant set to DTI measurements on articular cartilage in the hope that it will aid in understanding the nature of the collagen architecture which might exhibit a depth-dependence.  Our initial investigations suggest that our invariants show there is a difference in the structure between the superficial zone (oblateness) and the deep zone (prolateness) of articular cartilage. We are currently in the process of preparing this data for publication.    

What also remains to be verified is the claim that the curvilinear invariants possess better noise tolerance for approximately isotropic tensors. This could be achieved, for example, via the statistics of computer simulations of noisy tensors.

\section*{Conclusions}
We have explored the cylindrical and spherical invariant sets of the diffusion tensor that were proposed by Ennis and Kindlmann~\cite{Ennis2006}, showing how they relate to the geometry of the eigenvalue space and providing an alternative method for the proof of their orthogonality (via the metric). We have also shown how a log-Euclidean set is closely related to the cylindrical set and how this and the other sets are consistent with various averaging schemes. 

A curvilinear invariant set $\{C_i\}$ was developed which has a different structure to the preceding three sets, making orthotropy more explicit, and which is expected to be more suitable for measuring low degrees of anisotropy and isotropy. In addition to the usual analysis of eigenvectors and fractional anisotropy from DTI studies on tissues such as articular cartilage, the curvilinear invariants may provide an improved assessment of morphology and function of cartilage and other biological tissues.




\begin{thebibliography}{99}

\bibitem{LeBihan2012} Le Bihan D, Johansen-Berg H (2012) Diffusion MRI at 25: Exploring brain tissue structure and function. Neuroimage 61: 324--341.
\bibitem{Sinha2011} Sinha U, Sinha S, Hodgson JA, Edgerton RV (2011) Human soleus muscle architecture at different ankle joint angles from magnetic resonance diffusion tensor imaging. Journal of Applied Physiology 110: 807--819.
\bibitem{Benson2011} Benson AP, Bernus O, Dierckx H, Gilbert SH, Greenwood JP, et al. (2011) Construction and validation of anisotropic and orthotropic ventricular geometries for quantitative predictive cardiac electrophysiology. Interface Focus 1: 101--116.
\bibitem{deVisser2008_1} de Visser SK, Crawford RW, Pope JM (2008) Structural adaptations in compressed articular cartilage measured by diffusion tensor imaging. Osteoarthritis and Cartilage 16: 83--89.
\bibitem{Filidoro2005} Filidoro L, Dietrich O, Weber J, Rauch E, Oerther T, et al. (2005) High-resolution diffusion tensor imaging of human patellar cartilage: Feasibility and preliminary findings. Magnetic Resonance in Medicine 53: 993--998.
\bibitem{Meder2006} Meder R, de Visser SK, Bowden JC, Bostrom T, Pope JM (2006) Diffusion tensor imaging of articular cartilage as a measure of tissue microstructure. Osteoarthritis and Cartilage 14: 875--881.
\bibitem{LeBihan2001} Le Bihan D, Mangin JF, Poupon C, Clark CA, Pappata S, et al. (2001) Diffusion tensor imaging: Concepts and applications. Journal of Magnetic Resonance Imaging 13: 534--546.
\bibitem{Basser1996} Basser PJ, Pierpaoli C (1996) Microstructural and physiological features of tissues elucidated by quantitative-diffusion-tensor MRI. Journal of Magnetic Resonance Series B 111: 209--219.
\bibitem{Batchelor2005} Batchelor PG, Moakher M, Atkinson D, Calamante F, Connelly A (2005) A rigorous framework for diffusion tensor calculus. Magnetic Resonance in Medicine 53: 221--225.
\bibitem{Fletcher2007} Fletcher PT, Joshi S (2007) Riemannian geometry for the statistical analysis of diffusion tensor data. Signal Processing 87: 250--262.
\bibitem{Bahn1999} Bahn MM (1999) Invariant and orthonormal scalar measures derived from magnetic resonance diffusion tensor imaging. Journal of Magnetic Resonance 141: 68--77.
\bibitem{Hasan2004} Hasan KM, Alexander AL, Narayana PA (2004) Does fractional anisotropy have better noise immunity characteristics than relative anisotropy in diffusion tensor MRI? An analytical approach. Magnetic Resonance in Medicine 51: 413--417.
\bibitem{Peled1998} Peled S, Gudbjartsson H, Westin CF, Kikinis R, Jolesz FA (1998) Magnetic resonance imaging shows orientation and asymmetry of white matter fiber tracts. Brain Research 780: 27--33.
\bibitem{Westin2002} Westin CF, Maier SE, Mamata H, Nabavi A, Jolesz FA, et al. (2002) Processing and visualization for diffusion tensor MRI. Medical Image Analysis 6: 93--108.
\bibitem{Basser2000} Basser PJ, Pajevic S (2000) Statistical artifacts in diffusion tensor MRI (DT-MRI) caused by background noise. Magnetic Resonance in Medicine 44: 41--50.
\bibitem{Pierpaoli1996} Pierpaoli C, Basser PJ (1996) Toward a quantitative assessment of diffusion anisotropy. Magnetic Resonance in Medicine 36: 893--906.
\bibitem{Ennis2006} Ennis DB, Kindlmann G (2006) Orthogonal tensor invariants and the analysis of diffusion tensor magnetic resonance images. Magnetic Resonance in Medicine 55: 136--146.
\bibitem{Criscione2000} Criscione JC, Humphrey JD, Douglas AS, Hunter WC (2000) An invariant basis for natural strain which yields orthogonal stress response terms in isotropic hyperelasticity. Journal of the Mechanics and Physics of Solids 48: 2445--2465.
\bibitem{Pasternak2010} Pasternak O, Sochen N, Basser PJ (2010) The effect of metric selection on the analysis of diffusion tensor MRI data. Neuroimage 49: 2190--2204.
\bibitem{Arsigny2006} Arsigny V, Fillard P, Pennec X, Ayache N (2006) Log-euclidean metrics for fast and simple calculus on diffusion tensors. Magnetic Resonance in Medicine 56: 411--421.
\bibitem{Jeffrey1986} Jeffrey A (1986). Mathematics for Engineers and Scientists. Third ed. Berkshire, England: Van Nostrand Reinhold (UK). pp. 444--455.

\end{thebibliography}
\end{document}